\documentclass[pra,twoside,showpacs,superscriptaddress,twocolumn]{revtex4}

\usepackage{latexsym}

\usepackage{graphics}

\begin{document}

\title{Optical implementation of the encoding of two qubits to a single qutrit}

\author{Lucie Bart{\accent23 u}\v{s}kov\'{a}}
\affiliation{Department of Optics, Palack\'y University,
     17.~listopadu 50, 772\,00 Olomouc, Czech~Republic}

\author{Anton\'{\i}n \v{C}ernoch}
\affiliation{Department of Optics, Palack\'y University,
     17.~listopadu 50, 772\,00 Olomouc, Czech~Republic}

\author{Radim Filip}
\affiliation{Department of Optics, Palack\'y University,
     17.~listopadu 50, 772\,00 Olomouc, Czech~Republic}
\affiliation{Institut f\"ur Optik, Information und
    Photonik, Max-Planck Forschungsgruppe, Universit\"at
    Erlangen-N\"urnberg, G\"unther-Scharowsky str.~1, 91058,
    Erlangen, Germany}

\author{Jarom{\'\i}r Fiur\'{a}\v{s}ek}
\affiliation{Department of Optics, Palack\'y University,
     17.~listopadu 50, 772\,00 Olomouc, Czech~Republic}

\author{Jan Soubusta}
\affiliation{Joint Laboratory of Optics of Palack\'{y} University and
     Institute of Physics of Academy of Sciences of the Czech Republic,
     17. listopadu 50A, 779\,07 Olomouc, Czech Republic}

\author{Miloslav Du\v{s}ek}
\affiliation{Department of Optics, Palack\'y University,
     17.~listopadu 50, 772\,00 Olomouc, Czech~Republic}

\date{\today}

\begin{abstract}

We have devised an optical scheme for the recently proposed
protocol for encoding two qubits into one qutrit. In this
protocol, Alice encodes an arbitrary pure product state of two
qubits into a state of one qutrit. Bob can then restore error-free
any of the two encoded qubit states but not both of them
simultaneously. We have successfully realized this scheme
experimentally using spatial-mode encoding. Each qubit (qutrit)
was represented by a single photon that could propagate through
two (three) separate fibers. We theoretically propose two
generalizations of the original protocol. We have found a
probabilistic operation that enables to retrieve both qubits
simultaneously with the average fidelity above 90\,\% and we have
proposed extension of the original encoding transformation to
encode $N$ qubits into one $(N+1)$-dimensional system.

\end{abstract}


\maketitle

%%%%%%%%%%%%%%%%%%%%%%%%%%%%%%%%%%%%%%%%%%%%%%%%%%%%%

\section{Introduction}

A full description of the most elementary two-level quantum system (a
qubit)  in a pure state requires two real numbers -- the angular
coordinates of the point on the surface of the Bloch sphere which
specifies the state. Although an infinite amount of classical
information is required for its full characterization, the qubit
can be used to transmit only a finite amount of classical
information -- a single bit at most. This is an example of the
celebrated Holevo bound \cite{Holevo73} which states that no more
than $\log_2 d$ bits of information could be extracted from a
single copy of a $d$-level quantum system. The deep reason for
this lies in the process of quantum measurement which cannot
perfectly discriminate non-orthogonal quantum states and
consequently the optimal strategy to encode information is to
prepare the system in one out of $d$ orthogonal states.

Going beyond qubits, a qutrit (three level system) in a pure state
is specified by four real numbers (up to an irrelevant overall
phase). This is the same number of parameters which is necessary
to specify two uncorrelated qubits in a pure product state. Based
on this observation, Grudka and Wojcik (GW) investigated whether a
product state of two qubits could be somehow encoded onto a single
qutrit \cite{Grudka03}. Interestingly, they showed that this is
indeed possible, at least to certain extent. In particular they
proposed an encoding and decoding strategy which allows perfectly
extract either the first or the second qubit from the qutrit with
average probability $2/3$.

In the present paper we report on the experimental demonstration
of the GW protocol for optical qubits represented by single
photons propagating in two different optical fibers. This way of
encoding has been used quite rarely so far but it is very suitable
for our present purpose. In particular, it allows for very natural transition
from qubits to a qutrit encoded into a path of a single photon
which could propagate in three different optical fibers. This
approach might be thus advantageous in all situations where one
wishes to exploit higher-dimensional Hilbert spaces.

In addition to the experimental realization of the original GW
protocol, we have also extended it in several ways. First, we
have theoretically demonstrated that it is possible, with a certain
probability, to extract from the qutrit \emph{both} qubits
simultaneously. Of course, this decoding is imperfect but the
average fidelity of the decoded qubits, $F\approx 0.9023$, is
surprisingly high. We have also shown that the  protocol can be
easily generalized for several qubits or even qudits.

The rest of the paper is organized as follows. In Section II we briefly review the
GW protocol and present the optimal procedure for probabilistic simultaneous 
decoding of both qubits. We also describe here the extension of the protocol 
to more than two qubits and to qudits. The experimental
setup is described in Sec. III and the experimental results are presented in
Sec. IV. The paper ends with a brief summary and conclusions in Sec. V. 

\section{Theory}

Let us begin by briefly reviewing the GW protocol. Consider two
qubits labeled $1$ and $2$ prepared in pure states,
\begin{equation}
|\psi_1\rangle= \alpha_1|0\rangle_1+\beta_1|1\rangle_1, \qquad
|\psi_2\rangle= \alpha_2|0\rangle_2+\beta_2|1\rangle_2.
\end{equation}
Let $|0\rangle,~|1\rangle,~|2\rangle$ be the basis in the Hilbert
space of the qutrit into which we will  encode the qubits. The
probabilistic encoding operation then consists of the following
mapping,
\begin{eqnarray}
|0\rangle_1|0\rangle_2 &\rightarrow& |0\rangle, \nonumber \\
|0\rangle_1|1\rangle_2 &\rightarrow& |1\rangle, \nonumber \\
|1\rangle_1|1\rangle_2 &\rightarrow& |2\rangle.
\label{encoding}
\end{eqnarray}
Note that the state $|1\rangle_1|0\rangle_2$ is filtered out in
the mapping which is necessary in order to accommodate the two
qubits into a single qutrit. Consequently, the normalized state of
the qutrit after the encoding reads
\begin{equation}
|\Psi\rangle=
\frac{1}{\sqrt{\mathcal{N}}}(\alpha_1\alpha_2|0\rangle+
\alpha_1\beta_2|1\rangle+\beta_1\beta_2|2\rangle),
\label{qutrit}
\end{equation}
where $\mathcal{N}=1-|\beta_1|^2|\alpha_2|^2$ is the probability
of successful encoding. In the decoding procedure, the qutrit is
projected onto a two-dimensional subspace. The projectors
corresponding to the decoding of the first or the second qubit
read:
\begin{eqnarray}
&\Pi_{1+}= |1\rangle \langle 1|+|2\rangle \langle 2|, \qquad
&\Pi_{1-}=|0\rangle \langle 0|,  \nonumber \\
&\Pi_{2+}= |0\rangle \langle 0|+|1\rangle \langle 1|, \qquad
&\Pi_{2-}=|2\rangle \langle 2 |.
\end{eqnarray}
Projection onto $\Pi_{j+}$ indicates successful decoding of the
$j$th qubit while $\Pi_{j-}$ signals failure. Considering the
decoding of the first qubit, the resulting state is given by
\begin{equation}
|\psi_{\mathrm{out}}\rangle = \Pi_{1+}|\Psi\rangle =
\frac{1}{\sqrt{\mathcal{N}}}\beta_2(\alpha_1|1\rangle+\beta_1|2\rangle),
\end{equation}
hence $|\psi_{\mathrm{out}}\rangle\propto |\psi_1\rangle$ and the
decoding is perfect. The probability of the successful decoding of
the first qubit is $|\beta_2|^2/\mathcal{N}$. Assuming uniform
distribution of the qubits on the surface of the Bloch sphere, the
average probability of successful encoding and decoding is $1/2$. 
The same holds for the extraction of the second qubit.

The above described procedure  allows to perfectly extract one
qubit from the qutrit. We have investigated also an alternative
decoding strategy, where both qubits are retrieved simultaneously.
This unavoidably introduces some noise. Let $\rho_j$ denote the
(generally mixed) state of $j$th retrieved qubit, then we can
quantify the quality of the decoding procedure by the fidelities
$F_1 = \overline{F_1(\psi_1)}$ and $F_2 = \overline{F_2(\psi_2)}$
averaged over all possible input states, where $F_1(\psi_1)=
\langle \psi_1|\rho_1|\psi_1\rangle$ and $F_2(\psi_2)= \langle
\psi_2|\rho_2|\psi_2\rangle$. We have concentrated on a symmetric
retrieval where $F_1=F_2=F$ and with the help of the techniques
introduced in Ref. \cite{Fiurasek04} we have determined the
optimal probabilistic decoding operation which maximizes the
average fidelity $F$. This operation explicitly reads
\begin{eqnarray}
|0\rangle &\rightarrow & \frac{1}{\sqrt{2}}|0\rangle_1|0\rangle_2,  \nonumber \\
|1\rangle &\rightarrow & |0\rangle_1|1\rangle_2,  \nonumber \\
|2\rangle &\rightarrow & \frac{1}{\sqrt{2}}|1\rangle_1|1\rangle_2.
\label{jointretrieval}
\end{eqnarray}
and the corresponding average fidelity is
\begin{equation}
F=\frac{4+\sqrt{2}}{6} \approx 0.9024.
\end{equation}
Note that the operation (\ref{jointretrieval}) is not a direct
inversion of the encoding transformation (\ref{encoding}) since it
involves the prefactors $1/\sqrt{2}$. The average probability of
success of encoding  and decoding operations (\ref{encoding})
and (\ref{jointretrieval}) reads $P=1/2$. Note also that the
procedure is not covariant and various states are encoded and
decoded with different probabilities and fidelities. For a given
fixed state of the first qubit
$|\psi_1\rangle=\cos\frac{\vartheta}{2}|0\rangle_1+e^{i\varphi}
\sin\frac{\vartheta}{2}|1\rangle_1$ the probability of success of
the joint encoding/decoding procedure averaged over all possible
states of the second qubit is given by
\begin{equation}
P_1(\vartheta)=\frac{1}{4}+\frac{1}{2}\cos^2\frac{\vartheta}{2},
\end{equation}
and the corresponding normalized fidelity of the retrieved state
reads
\begin{equation}
F_1(\vartheta)=
\frac{1+2\cos^4\frac{\vartheta}{2}+\frac{\sqrt{2}-1}{2}\sin^2\vartheta}{1+2\cos^2\frac{\vartheta}{2}}.
\label{F1}
\end{equation}
Note that $P_1$ and $F_1$ do not depend on the phase $\varphi$.

The scheme can be also extended to probabilistic encoding and decoding
of $N$ qubits
$|\psi_j\rangle=\alpha_j|0\rangle_j+\beta_j|1\rangle_j$,
$j=1,\ldots,N$ into a $(N+1)$-dimensional state. The encoding
strategy is a straightforward generalization of the two-qubit
procedure (\ref{encoding}),
\begin{equation}
\bigotimes_{k=1}^j|1\rangle_k \bigotimes_{l=j+1}^N |0\rangle_l
\rightarrow |j\rangle, \qquad j=0,\ldots,N.
\end{equation}
This transformation produces the following state of the
$(N+1)$-dimensional system,
\begin{equation}
|\Psi_{N+1}\rangle= \sum_{j=0}^N \prod_{k=1}^j \beta_k \!
\prod_{l=j+1}^N \alpha_{l} \,|j\rangle.
\end{equation}
The POVM which decodes $n$-th qubit has the form
\begin{eqnarray}
\Pi_{n+} &=& |n-1\rangle \langle n-1|+|n\rangle \langle n|,
\nonumber \\
\qquad \Pi_{n-} &=& I-\Pi_{n+},
\end{eqnarray}
where $I$ denotes the identity operator.

The protocol can be further generalized to qudits. If we possess
$N$ qudit states then we can encode them to one
$[N(d-1)+1]$-dimensional system in such a way that an arbitrary
single qudit can be perfectly extracted from that state with a
certain probability. The principle is the same as in the previous
cases. To illustrate it, let us consider encoding of two qutrits
\begin{eqnarray}\label{qutr}
|\phi_1\rangle&=&\alpha_1|0\rangle_1+\beta_1|1\rangle_1+\gamma_1|2\rangle_1,\nonumber\\
|\phi_2\rangle&=&\alpha_2|0\rangle_2+\beta_2|1\rangle_2+\gamma_2|2\rangle_2,
\end{eqnarray}
into the state of a five-dimensional system. The encoding strategy
can be expressed as follows,
\begin{eqnarray}
|0\rangle_1|j\rangle_2 &\rightarrow& |j\rangle,
\quad j = 0, 1, 2,  \nonumber \\
|1\rangle_1|2\rangle_2 &\rightarrow& |3\rangle, \nonumber \\
|2\rangle_1|2\rangle_2 &\rightarrow& |4\rangle.
\end{eqnarray}

This results in a five-dimensional state which carries both
qutrits,
\begin{eqnarray}
|\Phi\rangle \propto \alpha_1\alpha_2 |0\rangle+\alpha_1\beta_2
|1\rangle
+\alpha_1\gamma_2 |2\rangle+\nonumber\\
\beta_1\gamma_2 |3\rangle+\gamma_1\gamma_2 |4\rangle.
\label{quditencoded}
\end{eqnarray}
The decoding POVM elements can be easily inferred from  the
structure of the state (\ref{quditencoded}) and we obtain,
\begin{equation}
\Pi_{1+}= \sum_{j=2}^4 |j\rangle \langle j|, \qquad \Pi_{2+}=
\sum_{j=0}^2 |j\rangle \langle j|.
\end{equation}

\section{Description of the experiment}

\begin{figure}
  \begin{center}
    %\smallskip
  \resizebox{\hsize}{!}{\includegraphics*{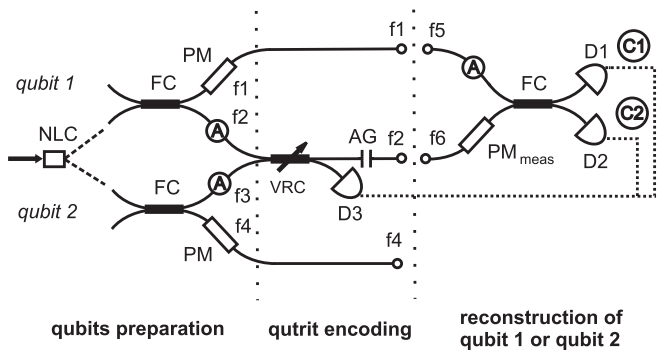}}
    %\smallskip
  \end{center}
  \caption{Setup of the experiment. \textsf{NLC} denotes a nonlinear crystal,
           \textsf{A} an attenuator, \textsf{FC} a fiber coupler, \textsf{PM}
           a phase modulator, \textsf{AG} an adjustable air gap, \textsf{VRC}
           a variable-ratio coupler, and \textsf{D1}-\textsf{D3} denote detectors.}
  \label{setup}
\end{figure}

Our experimental setup is shown in Fig.~\ref{setup}. The
experiment is based on the interplay of the second-order and
fourth-order interference. To prepare the two qubits we utilize
photon pairs generated by the process of type-I spontaneous
parametric down conversion (SPDC) in a 10-mm-long LiIO$_3$
nonlinear crystal (NLC) pumped by a krypton-ion cw laser
(413.1\,nm, 180\,mW). Down-converted beams filtered by cut-off
filters and circular apertures are coupled into single-mode
optical fibers. The photons in each generated pair are tightly
time correlated. Each of the photons enters a fiber coupler (FC)
and splits into two channels. The first qubit is represented by 
a single photon in fibers $f1$ and $f2$ while the second qubit 
corresponds to a single photon in fibers $f3$ and $f4$. 
The final state of each qubit is
determined by the adjustment of intensity ratio and phase shift
between these channels. To control the intensity ratio we use a
balanced fiber coupler  followed by an attenuator (A) in one of the
output arms (instead of a fiber coupler with a variable splitting
ratio). I.e., we use a conditional preparation procedure. The
phase difference is set by a phase modulator (PM). Equatorial
states of qubits are realized simply by changing just the phase
difference whereas the losses in both fibers are balanced.

%%%%%%%%%%%%%%%%%%%%%%%%%%%%%%%%%%%%%%%%%%%%%%%%%%

%%%%%%%%%%%%%%%%%%%%%%%%%%%%%%%%%%%%%%%%%%%%%%%%%%

The compression of two qubits into one qutrit is realized by a
variable-ratio coupler (VRC) and a subsequent measurement. This
operation is probabilistic and it succeeds only if detector D3
fires. VRC mixes modes $f2$ and $f3$. If a signal on detector D3
is detected there cannot be more then one photon in the remaining
fibers $f1$, $f2$, $f4$. These three modes constitute a qutrit. If
the described operation is successful the (non-normalized) state
of the qutrit reads
\begin{eqnarray}
\sqrt{T}\,\alpha_{1}\alpha_{2}
|\mathtt{001}\rangle_{f1f2f4}+(R-T)\,
\alpha_{1}\beta_{2} |\mathtt{010}\rangle_{f1f2f4}+ && \nonumber \\
\sqrt{R}\,\beta_{1}\beta_{2} |\mathtt{100}\rangle_{f1f2f4}, &&
\end{eqnarray}
where $R$ denotes the reflectance and $T=1-R$ the transmittance
of VRC ($R$ applies between ports $f2$-$f2$ and $f3$-$f3$ whereas
$T$ between $f2$-$f3$ and $f3$-$f2$),
$|\mathtt{001}\rangle_{f1f2f4}$ denotes the state with one photon
in mode $f4$, etc. To acquire the state of the qutrit in the form
given by Eq.~(\ref{qutrit}), suitable for the reverse decoding,
a filtration procedure represented by additional attenuations
has to be applied. The damping factor $\eta_{1}=(R-T)^{2}/R$ must
be applied in mode $f1$, whereas $\eta_{4}=(R-T)^{2}/T$ in mode
$f4$. Thus the probability of success of the whole encoding
transformation (including the filtration) is
$P=(T-R)^{2}(1-|\beta_{1}|^{2} |\alpha_{2}|^{2})$. Now we want to
find such a splitting ratio of VRC which maximizes this
probability. Of course, it is restricted by conditions $\eta_{1},
\eta_{4}\leq 1$. The optimal value of the splitting ratio is
$R=1/4, T=3/4$. Corresponding damping factors read $\eta_{1}=1$
(i.e., no attenuation) and $\eta_{4}=1/3$ (in the experiment $\eta_4$ is
set by increasing the attenuation of the attenuator in fiber
$f5$). 

We can thus see that an unbalanced coupler lies at the heart of 
the encoding procedure. 
In this context it is worth noting that unbalanced beam
splitters find several applications in optical quantum information processing
ranging from the realization of the quantum logic 
gates \cite{Langford05,Kiesel05,Okamoto05} to optimal universal
\cite{Zhao05,Filip04} 
and phase-covariant \cite{Fiurasek03} cloning of single-photon states.

From the created qutrit one can recover error-free either
of the two original qubits but not both of them. Choosing 
fibers $f1$ and $f2$ one can decode qubit~1, whereas selecting
$f2$ and $f4$ qubit 2 can be obtained. This procedure is also
probabilistic as there is nonzero probability that the photon is
in the remaining fiber. To check the states of reconstructed
qubits we use an interferometric measurement. Connecting fiber
$f1$ with $f5$ and $f2$ with $f6$ the reconstructed qubit~1 is
verified whereas connecting $f2$-$f6$ and $f4$-$f5$ qubit~2 can be
checked. Of course, only the cases when the encoding procedure
successfully occurred are taken into account.

\begin{figure}
  \begin{center}
    %\smallskip
  \resizebox{\hsize}{!}{\includegraphics*{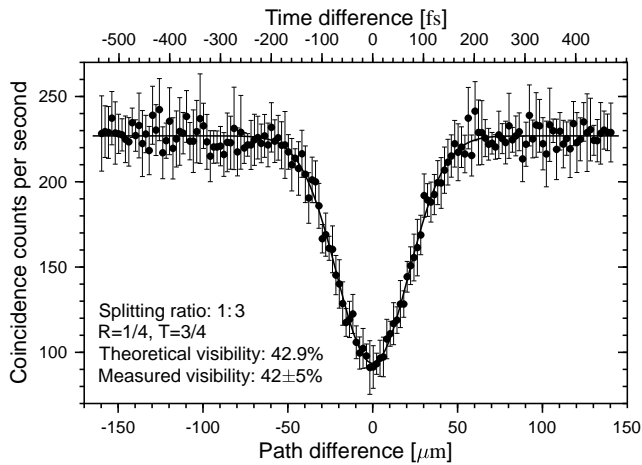}}
    %\smallskip
  \end{center}
  \caption{Hong-Ou-Mandel dip for the splitting ratio 25:75.
        The full line shows a Gaussian fit.}
  \label{dip}
\end{figure}

The whole process, including encoding and decoding of information,
depends on the quality of the fourth- and second-order
interference. Before starting the measurement it is necessary to
adjust the Hong-Ou-Mandel (HOM) interferometer \cite{HOM} formed
by VRC. First we set the VRC splitting ratio to 50:50 and adjusted
the precise time overlap of the two photons at VRC and tuned their
polarizations (this was done by mechanical fiber polarization
controllers not shown in the scheme). The visibility of HOM dip
was about $98\,\%$. Then we changed the VRC splitting ratio to
25:75. Fig.~\ref{dip} shows HOM dip measured with this splitting
ratio. During this measurement the attenuator in fiber $f5$ was
closed and the coincidences between detectors D1 and D3 and
between D2 and D3 were counted. In the graph the sum of these two
coincidence counts is plotted. Each point was averaged over eight
one-second measurements. For comparison the theoretical value of
visibility is $42.9\,\%$.

%%%%%%%%%%%%%%%%%%%%%%%%%%%%%%%%%%%%%%%%%%%%%%%%%%

%%%%%%%%%%%%%%%%%%%%%%%%%%%%%%%%%%%%%%%%%%%%%%%%%

The fiber-based Mach-Zehnder (MZ) interferometer was adjusted by
the following procedure: Only one beam from the nonlinear crystal
was used, the other one was blocked. To set the same optical
lengths of the arms of the interferometer an adjustable air-gap
(AG) was employed. Its precision is about $0.1\,\mu$m (the precise
setting of the phase difference was then done by an
electro-optical phase modulator $\mbox{PM}_{\mathrm{meas}}$).
First, losses in both arms of the interferometer were balanced
and the polarization states in  both arms were aligned (it was
done by mechanical fiber polarization controllers that are not
shown in the scheme). In this setting we reached visibilities
above $97\,\%$. Then we unbalanced the losses to take into account
the reflectivity of VRC ($R=1/4$) and the damping factors
$\eta_{1}$ or $\eta_{4}$. Then we opened both inputs again and let
both photons from each pair come into the system.

Fluctuations of temperature and temperature gradients cause the
changes of refraction indices of fibers. This is the reason for
the substantial instability of the interference pattern. Therefore
the interferometer must be thermally isolated (we use polystyrene
boxes). However, this is not sufficient -- the phase difference between
the arms of the interferometer still drifts in time by about
$\pi/1000$ per second on average. The environmental perturbations
may further be reduced by means of active stabilization. In the
experiment five-second measurement blocks are alternated by
stabilization cycles, each taking also about 5 seconds in average.
In each stabilization cycle the value of the phase drift is
estimated and it is compensated by means of a phase modulator
$\mbox{PM}_{\mathrm{meas}}$. During the stabilization only one
beam from the crystal is allowed to enter the system, the other
one is blocked. Combination of the passive and active methods of
stabilization gives very good results.

The experiment itself starts with the setting of input qubit
states by means of attenuators and phase modulators in fibers
$f1$, $f2$, $f3$ and $f4$. Then the proper measurement basis for
the verification of the state of a reconstructed qubit is set by
adjusting the attenuator and phase modulator in the measurement
part of the setup (fibers $f5$ and $f6$). The measurement basis
consists of the original state of the corresponding qubit and of
the state orthogonal to it. Then the coincidences between
detectors D1 and D3 and coincidences between D2 and D3 are
counted. We use Perkin-Elmer single-photon counting modules
(employing silicon avalanche photodiodes with quantum efficiency
$\eta \approx 60\,\%$ and dark counts about $50\,\mathrm{s}^{-1}$)
and coincidence electronics based on time-to-amplitude convertors
and single-channel analyzers with a two-nanosecond coincidence
window. In the ideal case the coincidences should be detected only
between detectors D1 and D3 -- these events correspond to the
detection of the original state of the qubit; we denote the
corresponding coincidence rate as $C^{+}$. Coincidences between
detectors D2 and D3 represent erroneous detections and we denote
the corresponding coincidence rate as $C^{-}$. Fidelity of the
reconstructed qubit state is then
\begin{equation}
F=\frac{C^{+}}{C^{+}+C^{-}}.
\end{equation}

To realize experimentally the encoding of $N$ qubits into one
$(N+1)$-dimensional system  one can repeatedly use
the unit depicted in Fig.~\ref{setup} (i.e., add other
Mach-Zehnder interferometers interconnected by variable-ratio
couplers). However, in such a case one would need $N$
time-correlated photons.

%%%%%%%%%%%%%%%%%%%%%%%%%%%%%%%%%%%%%%%%%%%%%%%%%%

\begin{figure}
  \begin{center}
    %\smallskip
  \resizebox{\hsize}{!}{\includegraphics*{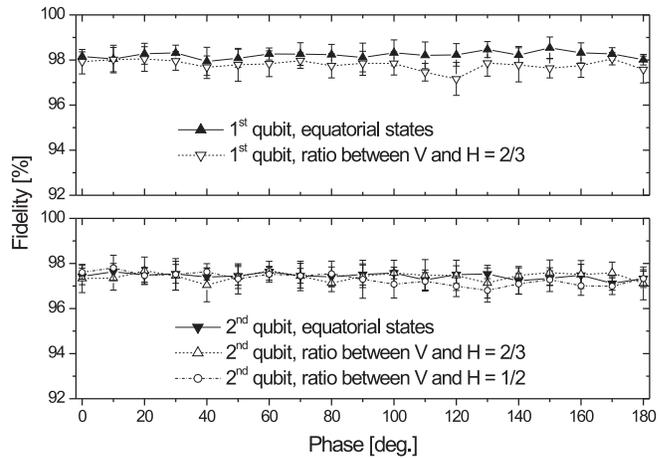}}
    %\smallskip
  \end{center}
  \caption{Observed fidelities of reconstructed qubit states
           for various input states.}
  \label{fid}
\end{figure}

%%%%%%%%%%%%%%%%%%%%%%%%%%%%%%%%%%%%%%%%%%%%%%%%%

\section{Experimental results}

We encoded and decoded the states of qubits,
\begin{eqnarray}
 |\psi\rangle &=&
 \left[ \cos{\frac{\vartheta_{1}}{2}} |\mathtt{01}\rangle_{f1f2} +
  e^{i\varphi_{1}} \sin{\frac{\vartheta_{1}}{2}} |\mathtt{10}\rangle_{f1f2}\right]
 \otimes   \nonumber\\
 &&\left[ \cos{\frac{\vartheta_{2}}{2}} |\mathtt{01}\rangle_{f3f4} +
  e^{i\varphi_{2}}\sin{\frac{\vartheta_{2}}{2}}
  |\mathtt{10}\rangle_{f3f4}
 \right],
\end{eqnarray}
with the following parameters: For qubit~1 we checked 38 different
states:
\begin{itemize}
\item $\vartheta_{1} = 90^\circ$
  [i.e., $|\alpha_1|^2 : |\beta_1|^2 = 1:1 $],
  \\$\varphi_{1}=0^\circ, 10^\circ,\dots, 180^\circ$;
  \\ $\vartheta_{2}$, $\varphi_{2}$ were arbitrary;
\item $\vartheta_{1} = 78.46^\circ$,
  [i.e., $|\alpha_1|^2 : |\beta_1|^2 = 3:2$],
  \\$\varphi_{1}=0^\circ, 10^\circ,\dots, 180^\circ$;
  \\ $\vartheta_{2}$, $\varphi_{2}$ were arbitrary.
\end{itemize}
For qubit~2 we checked even 57 different states:
\begin{itemize}
\item $\vartheta_{2} = 90^\circ$,
  [i.e., $|\alpha_2|^2 : |\beta_2|^2 = 1:1$],
  \\$\varphi_{2}=0^\circ, 10^\circ,\dots, 180^\circ$;
  \\$\vartheta_{1}$, $\varphi_{1}$ were arbitrary;
\item $\vartheta_{2} = 78.46^\circ$,
  [i.e., $|\alpha_2|^2 : |\beta_2|^2 = 3:2$],
  \\$\varphi_{2}=0^\circ, 10^\circ,\dots, 180^\circ$;
  \\ $\vartheta_{1}$, $\varphi_{1}$ were arbitrary;
\item $\vartheta_{2} = 70.53^\circ$,
  [i.e., $|\alpha_2|^2 : |\beta_2|^2 = 2:1$],
  \\$\varphi_{2}=0^\circ, 10^\circ,\dots, 180^\circ$;
  \\ $\vartheta_{1}$, $\varphi_{1}$ were arbitrary.
\end{itemize}
When we verified the state of the reconstructed qubit~1 the
qubit~2 was prepared in an arbitrary state and vice versa.

Observed fidelities of reconstructed qubit states are shown in
Fig.~\ref{fid}. Each point was calculated as an average from 10
five-second measurements (we detected hundreds of coincidences
$C^{+}$ per second -- the exact numbers varied according to
attenuations). Fidelities exhibit values around 98\,\% and are
independent on the qubit states (i.e., constant within the 
 statistical errors). The deviations from the ideal 100\,\%
fidelity are caused by misalignments, inaccuracies in polarization
settings, limited precision of parameter setting, and a phase
drift (during the measurement period) in the MZ interferometer.

\section{Conclusions}

We have experimentally realized transformation for the encoding of
two single-photon qubits into one qutrit and demonstrated that it is possible to
restore (probabilistically but error-free) any of the two encoded
qubit states. The principle of the experiment is based on the interplay of the
second-order and fourth-order interference.
We have reached fidelities around 98\,\%. The
deviations from the ideal 100\,\% are caused by imperfections of
the experimental setup. We have employed encoding to spatial modes
which has been used quite rarely so far but it was
very suitable for our purpose, since it allows for very natural
transition from qubits to a qutrit encoded into path of a single
photon propagating in three different optical fibers. Note that, in principle, 
the scheme could also work with other encodings that admit
higher dimensional Hilbert spaces such as time-bin \cite{Marcikic04,Stucki05} 
or orbital angular momentum \cite{Mair01,Molina04} encodings.

Further, we have proposed some generalizations of the encoding and
decoding protocol. We have found a probabilistic operation that
allows  to retrieve  both qubits simultaneously with the average
fidelity above 90\,\%. We have also proposed an extension of the
original encoding transformation to encode $N$ $d$-dimensional
systems into one $[N(d-1)+1]$-dimensional system. Implementation
of the encoding of $N$ qubits into one $(N+1)$-dimensional system
can be done using repeatedly Mach-Zehnder interferometers
interconnected by variable-ratio couplers.

% (however, in such a case
%one needs $N$ time-correlated photons). 

%%%%%%%%%%%%%%%%%%%%%%%%%%%%%%%%%%%%%%%%%%%%%%%%%%%%%%%%%%%%%%%%%%%

\begin{acknowledgments}

This research was supported by the projects MSM6198959213 and
LC06007 of the Ministry of Education of the Czech Republic, by GA\v{C}R
(202/03/D239) and by the SECOQC project of the EC (IST-2002-506813).

\end{acknowledgments}

%%%%%%%%%%%%%%%%%%%%%%%%%%%%%%%%%%%%%%%%%%%%%%%%%%%%%%%%%%%%%%%%%%%


\begin{thebibliography}{99}


\bibitem{Holevo73}
A.\,S.~Holevo, Probl.\ Pereda.\ Inf.\ \textbf{9}, 3 (1973)
[Probl.\ Inf.\ Transm.\ \textbf{9}, 110 (1973)].

\bibitem{Grudka03}
A.~Grudka and A.~Wojcik, Phys.\ Lett.\ A \textbf{314}, 350 (2003).

\bibitem{Fiurasek04}
J.~Fiur\'{a}\v{s}ek, Phys.\ Rev.\ A \textbf{70}, 032308 (2004).




%Demonstration of a Simple Entangling Optical Gate and Its Use in Bell-State Analysis
\bibitem{Langford05}
N. K. Langford, T. J. Weinhold, R. Prevedel, K. J. Resch, A. Gilchrist, J. L. O'Brien, 
G. J. Pryde, and A. G. White, 
Phys. Rev. Lett. \textbf{95}, 210504  (2005).

\bibitem{Kiesel05}
%Linear Optics Controlled-Phase Gate Made Simple
N. Kiesel, Ch. Schmid, U. Weber, R. Ursin, and H. Weinfurter,
Phys. Rev. Lett. \textbf{95}, 210505  (2005).

\bibitem{Okamoto05}
% Demonstration of an Optical Quantum Controlled-NOT Gate without Path Interference
R. Okamoto, H. F. Hofmann, S. Takeuchi, and K. Sasaki,
Phys. Rev. Lett. \textbf{95}, 210506  (2005).

\bibitem{Zhao05}
Z. Zhao, A.-N. Zhang, X.-Q. Zhou, Y.-A. Chen, C.-Y. Lu, A. Karlsson, 
and J.-W. Pan, Phys. Rev. Lett. \textbf{95}, 030502 (2005). 

\bibitem{Filip04}
R. Filip, Phys. Rev. A \textbf{69}, 052301 (2004).


\bibitem{Fiurasek03}
J. Fiur\'{a}\v{s}ek, Phys. Rev. A \textbf{67}, 052314 (2003).

\bibitem{HOM}
C.\,K.~Hong, Z.\,Y.~Ou, and L.~Mandel, Phys.\ Rev.\ Lett.\
\textbf{59}, 2044 (1987).


\bibitem{Marcikic04} 
I. Marcikic, H. de Riedmatten, W. Tittel, H. Zbinden, M. Legre, and N. Gisin,
Phys. Rev. Lett. \textbf{93}, 180502 (2004). 

\bibitem{Stucki05}
D. Stucki, H. Zbinden, and N. Gisin,
J. Mod. Opt. \textbf{52}, 2637 (2005). 


\bibitem{Mair01}
 A. Mair, A. Vaziri, G. Weihs, and A. Zeilinger, 
 Nature (London) \textbf{412}, 313 (2001).

\bibitem{Molina04}
G. Molina-Terriza, A. Vaziri, J. \v{R}eh\,{a}\v{c}ek, Z. Hradil, and A. Zeilinger,
Phys. Rev. Lett. \textbf{92}, 167903 (2004). 


\end{thebibliography}
\end{document}